\documentclass[twocolumn]{aastex63}
\usepackage{amsmath}
\usepackage{comment}
\usepackage[T1]{fontenc}
\usepackage{graphicx}
\usepackage{hyperref}
\usepackage{isotope}
\usepackage{longtable}
\usepackage[version=4]{mhchem}
\usepackage{multirow}
\usepackage{soul}
\usepackage{subfigure}
\usepackage{tablefootnote}
\usepackage{todonotes}
\usepackage{xcolor}

\newcommand{\rp}{\emph{r}-process}

\newcommand{\fref}[1]{Figure~\ref{#1}}

\newcommand{\sref}[1]{Section~\ref{#1}}

\begin{document}
%-----------------------------------------------------------------%
\title{Kilonova  Emissions from Neutron Star Merger Remnants: Implications for Nuclear Equation of State}

%-----------------------------------------------------------------%
\author[0000-0003-0031-1397]{Kelsey A. Lund}
\affiliation{Department of Physics, North Carolina State University, Raleigh, NC 27695, USA}
\affiliation{Center for Nonlinear Studies, Los Alamos National Laboratory, Los Alamos, NM 87545, USA}
\affiliation{Theoretical Division, Los Alamos National Laboratory, Los Alamos, NM 87544 USA}

\author[0000-0003-0427-3893]{Rahul Somasundaram}
\affiliation{Department of Physics, Syracuse University, Syracuse, New York 13244, USA}
\affiliation{Theoretical Division, Los Alamos National Laboratory, Los Alamos, NM 87544 USA}

\author[0000-0001-6811-6657]{Gail C. McLaughlin}
\affiliation{Department of Physics, North Carolina State University, Raleigh, NC 27695, USA}

\author[0000-0001-6432-7860]{Jonah M. Miller}
\affiliation{CCS-2, Los Alamos National Laboratory, Los Alamos, New Mexico 87545, USA}

\author[0000-0002-9950-9688]{Matthew R. Mumpower}
\affiliation{Theoretical Division, Los Alamos National Laboratory, Los Alamos, NM 87544 USA}
\affiliation{Center for Theoretical Astrophysics, Los Alamos National Laboratory, Los Alamos, NM 87544 USA}

\author[0000-0003-2656-6355]{Ingo Tews}
\affiliation{Theoretical Division, Los Alamos National Laboratory, Los Alamos, NM 87544 USA}
%-----------------------------------------------------------------%

\begin{abstract}
Multimessenger observations of binary neutron star mergers can provide valuable information on the nuclear equation of state (EOS).
Here, we investigate to which extent electromagnetic observations of the associated kilonovae allow us to place constraints on the EOS.
For this, we use state-of-the-art three-dimensional general-relativistic magnetohydrodynamics simulations and detailed nucleosynthesis modeling to connect properties of observed light curves to properties of the accretion disk, and hence, the EOS. 
Using our general approach, we use multimessenger observations of GW170817/AT2017gfo to study the impact of various sources of uncertainty on inferences of the EOS. 
We constrain the radius of a $\rm{1.4 M_\odot}$ neutron star to lie within $\rm{10.30\leq R_{1.4}\leq 13.0}$~km and the maximum mass to be $\rm{M_{TOV}\leq 3.06 M_\odot}$.\\
\end{abstract}


\section{Introduction}
The nuclear equation of state (EOS) describes the pressure of dense nuclear matter as a function of density, temperature, and composition.
Probing the dependence of the EOS on density and neutron-to-proton (isospin) asymmetry represents a grand challenge in nuclear physics given the difficulties associated with creating high densities and very asymmetric systems in terrestrial laboratory experiments \citep{Danielewicz2002,Russotto2016}.
Neutron stars, however, explore matter at high densities and isospin asymmetry, and hence, provide an excellent astrophysical laboratory for studying the EOS \citep{Lattimer2012}.
Explosive astrophysical events involving neutron stars are particularly important as they offer an additional avenue via which to probe dense nuclear matter under extreme conditions. 
Great effort is being dedicated toward building statistical frameworks for EOS inference from astronomical multimessenger observations, including binary neutron star mergers \citep[NSM;][]{Abbott2017a, Abbott2018,Bauswein2017, Coughlin2019, Miller2020, Radice2018c, Capano2020, Dietrich2020, Raaijmakers2020, Essick2021, Essick2021b, Ghosh2022, Huth2022, Pang2023, Takatsy2023, Zhu2023, Fan2024}.
These statistical models largely rely on piecing together different stages of the merger, making assumptions at each step.
For example, the nuclear EOS affects the behavior of neutron stars during the inspiral phase of an NSM \citep{Takami2014,Abbott2018,Most2019} as well as the properties of the postmerger system.
This system can generally  be characterized by an accretion disk surrounding a central remnant, either a heavy neutron star ($\rm{M \gtrsim 2 M_\odot}$) or a black hole \citep{Baumgarte2000, Kiuchi2012, Bauswein2013b, Lippuner2017, Metzger2018, Radice2018b, VanPutten2019, Ciolfi2020, Beniamini2021}.

The ejecta from this accretion disk are a promising site for the nucleosynthesis of the heaviest elements via the rapid neutron capture process (r-process), the decays of which power an electromagnetic transient.
Recent decades have seen immense efforts toward understanding the relation between the formation of the disk, its evolution, and the amount of material (especially \rp -producing material) that becomes unbound from the disk \citep{Ruffert1997, Popham1999, Shibata2007, Surman2008, Fernandez2013, Fernandez2014, Janiuk2014, Foucart2015, Just2015, Sekiguchi2015, Siegel2017, Fernandez2018, Miller2019, deHaas2023, Lund2024, Sprouse2024}, as well as its effectiveness (compared to tidal and shock-driven dynamical ejecta) at robustly producing \rp{} material.
The nuclear EOS plays a role in determining the distribution of material during and after the merger, affecting such quantities as the remnant disk mass and ejecta masses as well as the behavior of the late-time electromagnetic signal (the kilonova) that accompanies the merger event \citep{Radice2017, Radice2018, Abbott2018, Coughlin2018, Malik2018, Gamba2019, Kruger2020}. 
In particular, the mass of the disk ejecta is a key quantity involved in interpreting the kilonova signal attributed to the disk~\citep{Korobkin2021, Holmbeck2022, Ricigliano2024}. 

The NSM resulting in the combined electromagnetic and gravitational-wave observations from GW170817/AT2017gfo \citep[][and many more]{Abbott2017, Abbott2017a, Alexander2017, Cowperthwaite2017, Villar2017} remains the most closely scrutinized multimessenger event in recent years.
While many early works have used the inspiral, increasingly more works use both electromagnetic plus gravitational-wave signals to constrain the EOS \citep{Bauswein2017, Dietrich2017, Margalit2017, Radice2017, Wang2019, Breschi2021, Breschi2024, Pang2023}.

Here, we evaluate important physical considerations and potential degeneracies involved in several common steps in the inverse problem of using electromagnetic NSM observables to infer the EOS.
In \sref{sec:disk}, we analyze disk masses from numerical relativity (NR) simulations of NSMs published in the literature. From these, we introduce a novel fitting formula aimed at predicting the disk mass resulting from an NSM.
In \sref{sec:wind}, we make connections between the predicted disk masses with the mass of the disk ejecta, informed by 3D general-relativistic magnetohydrodynamics (3DGRMHD) simulations.
In \sref{sec:KNLC}, we connect the nuclear physics involved in the nucleosynthesis occurring in these ejecta to observable properties of the kilonova light curve.
In \sref{sec:EOSconnect}, we employ our disk mass formula to a set of chiral effective field theory ($\chi$EFT) informed nuclear EOSs, as described in \sref{sec:EOS}, for binaries consistent with the inferred properties of GW170817 in order to constrain the nuclear EOS using observations of AT2017gfo.
In \sref{sec:discussion}, we discuss our results.

\section{Postmerger Disk Mass}\label{sec:disk}

The nuclear EOS describes the properties of dense matter, which is a crucial input for understanding the behavior of neutron stars in explosive astrophysical events. 
It is a key input in NR simulations of NSMs as it plays a large part in determining the dynamics of the merger as well as the properties of the postmerger system. 
The properties of the system at the end of an NR simulation determine the initial conditions for GRMHD simulations of the remnant disk.
One of these properties is the disk mass.
The connection between an arbitrary binary and the remnant disk mass is often made via analytic formulae, informed by NR simulations. These relate an EOS-dependent quantity with the disk.

We discuss some of the proposed methods for using these data to analytically compute a remnant disk mass the Appendix, but here highlight that three major relations have been observed and are commonly used.
\citet{Radice2018} found a dependence of the final disk mass on the binary tidal deformability, $\tilde{\Lambda}$. 
\citet{Coughlin2019} highlighted a dependence on the threshold mass beyond which prompt collapse to a black hole occurs; \citet{Dietrich2020} refined this to include a dependence on the binary mass ratio, q. 
Finally, \citet{Kruger2020} found that the compactness of the lighter neutron star %$\rm{C_{light}}$
\footnote{The literature often uses subscripts (1,2) followed by a designation of each to either the lightest or heaviest NS in the binary. However, due to the lack of consistent designation of the smallest/largest component in the literature, throughout this work, we use subscripts ``light'' and \lq\lq heavy" to avoid any confusion.},
\begin{equation}
    \rm{C_{light} = \frac{GM_{light}}{c^2R_{light}}},
\end{equation}
in the binary was a good predictor of the resulting disk mass. 

\begin{figure}
    \centering
    \includegraphics[scale=0.65]{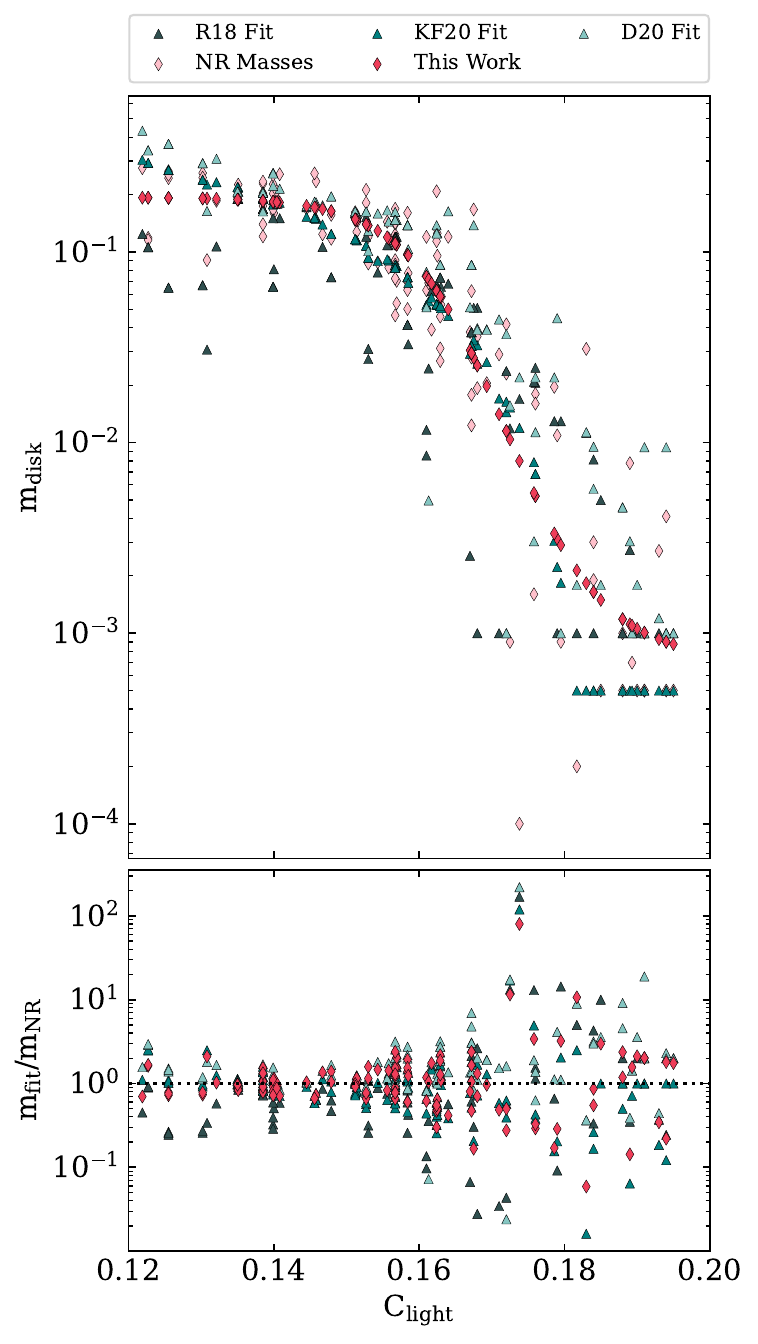}
    \caption{\textit{Top:} compilation of disk masses, as a function of the smallest NS compactness, $\rm{C_{\rm light}}$, for the 112 NR simulation points described in the main text (light pink diamonds).
    We show as triangles the results from the fitting formulae presented in R18 (dark blue), KF20 (teal), and D20 (light blue). The results from Equation~\eqref{eq:fit} are shown as dark pink diamonds. \textit{Bottom:} ratio between the masses obtained from the fit formulae to the NR data points.}
    \label{fig:Cfit}
\end{figure}

We compile disk masses from existing NR simulations, including those in \citet{Radice2018} and \citet{Kiuchi2019}, and those compiled by \citet{Camilletti2024}. 
The compilation from \citet{Camilletti2024} includes data from \citet{Nedora2019}, \citet{Perego2019}, \citet{Bernuzzi2020}, \citet{Endrizzi2020}, \citet{Nedora2021}, \citet{Cusinato2022}, \citet{Perego2022}, and \citet{Camilletti2022}.
This results in a total of 112 NR simulation data points from 11 sources; these are shown in Figure~\ref{fig:Cfit}.
Given this larger data set, we take the opportunity to reevaluate the aforementioned disk mass formulae.
We find that the dependence on $\rm{C_{light}}$ continues to yield a reasonably good fit, albeit with a different functional form than in \citet{Kruger2020}:
\begin{equation}\label{eq:fit}
    \rm{log_{10}\left ( m_{disk} \right ) = \alpha\ \tanh\left ( \beta\ C_{light} + \gamma \right ) + \delta}\,,
\end{equation}
with best-fit parameters $\alpha=-1.21$, $\beta=72.62$, $\gamma=-12.48$, and $\delta=-1.93$. 
We note that in determining this fit we do not consider non-EOS related dependences in the simulations, which might contribute to the spread in Figure~\ref{fig:Cfit}.

We show the predicted disk masses from our fit in the top panel of Figure~\ref{fig:Cfit} alongside those obtained from the fits of \citet[R18]{Radice2018}, \citet[KF20]{Kruger2020}, and \citet[D20]{Dietrich2020}. 
The bottom panel shows the ratio of the different disk mass predictions to the NR results. 
We find that our formulation provides a slightly better fit, with an overall RMS error of 0.042, compared to 0.065, 0.056, and 0.048 from R18, D20, and KF20, respectively.  
We also note that our fit has a simple functional form without an artificially imposed termination point or cutoff. A simple form is convenient given the ease with which the parameters can be adjusted to new data. 

%%%%%%%%%%%%%%%%%%%%%%%%%

\section{Disk to Ejecta Mass}\label{sec:wind}

The mass of the ejecta from the remnant accretion disk is an open question. Recent works have shown that enough \rp{} material can become unbound from these disks to account for the entire red component of the kilonova \citep{Siegel2017,Siegel2018}.
Motivated in part by these results, we focus \textit{solely} on the disk ejecta, and make the simplifying assumption that they alone are responsible for the entire red kilonova component, thus, exploring a limiting scenario of the effect of the \textit{total} merger ejecta on the kilonova. 
We point out, however, that the extent to which the dynamical versus disk component of the ejecta is responsible for the red kilonova remains an open question. 
We point the interested reader to efforts aimed at simulating the merger and postmerger phases self-consistently, such as the results presented in \citet{Kiuchi2023}.

Proceeding under the limiting-case assumptions we described above, for a given disk mass we need to determine the amount of material that is ejected. % from the disk. 
The most detailed evolution of material in the postmerger accretion disk is obtained via 3DGRMHD simulations, which combine the effects of magnetically driven turbulence, radiation transport, and neutrino interactions \citep{Gammie2003,Noble2006,Miller2019_nubh}. 
We discuss results from various 3DGRMHD post-NSM disk simulations and some of the differences across these different works, as well as the implications for the interpretation of their results.

\citet{Siegel2017,Siegel2018} presented the earliest detailed 3DGRMHD simulation of a remnant black hole accretion-disk system. 
An initial system consisting of a $0.03 \rm{M_\odot}$ torus surrounding a $3 \rm{M_\odot}$ black hole resulted in an ejecta mass of $\sim 0.2\rm{m_{disk}}$ after 381~ms, though the authors project an actual unbound mass fraction of $\sim 40\%$ based on the black hole accretion rate.
\citet{Fernandez2018} presented a simulation with similar initial conditions with an initial gas-to-magnetic pressure ratio of 100 instead of 200. 
The major result from this work was the evolution of the disk over 9.3~s and a resulting 40\% of the original disk becoming unbound. 
The authors concluded that by the end of this extended simulation time, the mass ejection is mostly concluded, and extending the simulation time would provide minimal returns.
Subsequent work by \citet{Christie2019} built on \citet{Fernandez2018} by evolving one weakly magnetized disk ($\beta = 850$\footnote{The parameter $\beta$ represents the ratio of gas-to-magnetic pressure and is commonly used in the context of magnetohydrodynamics simulations to quantify magnetization.}) and one disk with a strong ($\beta=5$) toroidal seed magnetic field, both with the same initial conditions.
Evolution of these two disks over $\sim 4$~s found ejecta mass fractions of 30\% and 27\%, respectively.

Detailed neutrino transport was incorporated into the work presented in \citet{Miller2019} and \citet{Sprouse2024}, which evolved the same weakly magnetized ($\rm{\beta=100}$), $\rm{M_{BH}\ (m_{disk}) = 2.58\ (0.12)\ M_\odot}$ system to 127~ms and 1.27~s, respectively.
The extended simulation time allowed for $\sim 30\%$ of the original disk to become unbound, with an indication (based on the mass ejection rate) that more mass could become unbound had the simulation run even longer.

\begin{figure*}[t]
    \centering
    \includegraphics[scale=0.6]{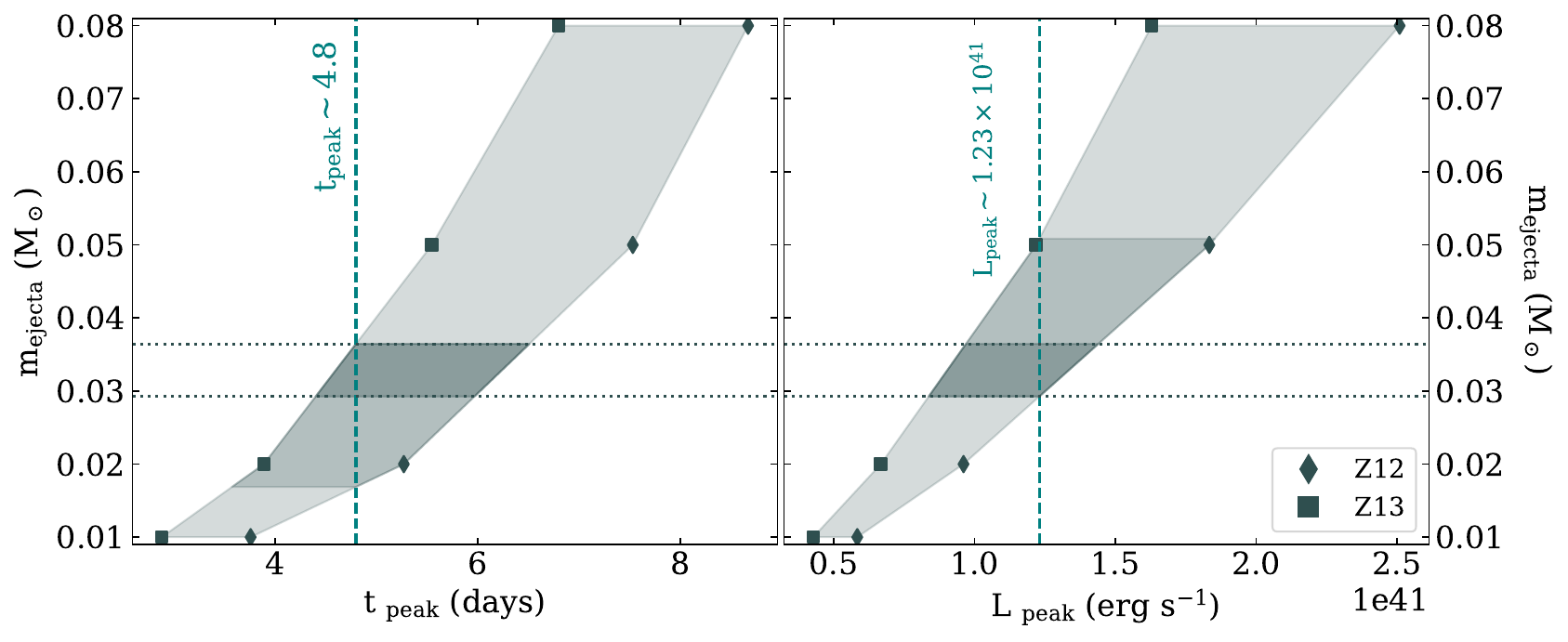}
    \caption{Inferred peak luminosity and times from the late-time, red component assuming a two-component model to explain the GW170817 electromagnetic signal (dashed vertical lines). Shaded regions highlight variation of these quantities with respect to ejecta mass from the Z12 and Z13 models of \citet{Zhu2021}. 
    The second darkest shaded region shows the region through which the peak time (left) and luminosity (right) overlap, while the darkest shaded region shows the values through which both of the inferred observed quantities overlap (also denoted by horizontal dotted lines).}
    \label{fig:Zmodels}
\end{figure*}

While these long-term simulations provide valuable insight into remnant accretion-disk mass ejection, their computational cost makes surveying different initial conditions while maintaining high-fidelity physics prohibitive. 
For example, the suite of simulations from \citet{Lund2024} also incorporates detailed neutrino transport (as in \citet{Miller2019}), but only tracks $\mathcal{O} \rm{(100~ms)}$ of evolution, albeit with different initial magnetic field strengths.
Similar to \citet{Christie2019}, these simulations indicate possible variation in the ejecta mass as well as the properties of the ejecta.
One main result was the larger ejecta mass resulting from stronger initial magnetic fields, from $< 2\%$ (for the same disk as in \citet{Miller2019} and \citet{Sprouse2024}) to more than 6\%.
It is not immediately clear if this increased mass ejection over the short simulation timescale would be sustained at later times.
We note that either way, differences in the timescale and geometry of mass (especially lanthanide) ejection stemming from different initial disk conditions could have more subtle impacts on the light curve. 
Based on the results of all these works, we proceed estimating that 30-40\% of the initial disk will become unbound, and that this range is reasonable to capture uncertainties from variations in the initial conditions of the disk.

\section{Kilonovae from Disk Ejecta}\label{sec:KNLC}

In order to relate ejecta mass to kilonova light-curve observables, we now turn to the electromagnetic observations from AT2017gfo and the wealth of observations and analyses from this event for two key observables: the peak luminosity of the ``red'' part of the kilonova and the time at which this peak occurs. 
The red component describes the behavior of the light curve after $\sim4$ days and can be largely attributed to ejecta from the postmerger system~\citep{Kasen2017, Waxman2019, Zhu2021}. 
We connect the properties of the red light curve and the disk ejecta mass following \citet{Zhu2021}, which explored the wide variety of nuclear physics uncertainties and their effect on kilonova light curves. 
We use their two models to describe ground-state binding energies of atomic nuclei (based on \citet{Duflo1995} and \citet{Kortelainen2012}) and corresponding linear combinations of parameterized, single-$\rm{Y_e}$ trajectories\footnote{`` Trajectories'' refers to the time evolution of the temperature and density, which is a key ingredient for nucleosynthesis calculations. These can be parameterized or informed by Lagrangian tracer particles from large-scale simulations} that were constructed to obtain a roughly solar final abundance pattern. The light curves were obtained following the procedure described in Section 4.2 of \citet{Zhu2021}.

The purpose of comparing these two models is to gauge the uncertainties from the unknown properties of nuclei far from stability with those from changing the ejecta mass. 
We note that the use of these two models likely underestimates the true uncertainty from nuclear physics as these models were constructed such that the resulting abundance pattern roughly matched the solar pattern; observations of AT2017gfo only indicate the production of lanthanides, with no direct proof that a solar pattern was produced. 
It is important to highlight that these models, though producing very similar patterns, undergo different nuclear heating histories, which result in different light-curve evolution. 
For each model, light curves were computed based on the nuclear heating and using ejecta masses of 0.01, 0.02, 0.05, and 0.08 $\rm{M_\odot}$, with the results shown in Figure~12(a) of \citet{Zhu2021}.
For each of the masses previously listed, we show the times at which the peak bolometric luminosities occurred, $t_{\rm peak}$, and the peak bolometric luminosities $L_{\rm peak}$ in the left and right panels of Figure~\ref{fig:Zmodels}, respectively. 
From the resulting bands, we are able to estimate an uncertainty in the properties of the kilonova originating from unknown nuclear physics.

This work highlights that a single ejecta mass can result in differences of 1-2 days in $t_{\rm peak}$. 
Similarly, $L_{\rm peak}$ for a given ejecta mass is subject to these same uncertainties, thus a single ejecta mass can be inferred from a variety of light-curve behaviors, which themselves are influenced by the unknown properties of nuclei far from stability. 
Thus, if one is attributing the late-time ($\gtrsim 1$ day) behavior of the kilonova to a disk wind, the interpretation of that late-time behavior depends to some extent on some combined assumption of both the nuclear heating history, the composition of the ejecta, and the ejecta mass itself. 

We show this by selecting a luminosity of $1.23\times10^{41}\rm{erg\ s^{-1}}$ occurring at roughly 4.8 days postmerger.
These are based on the two-component model shown in Fig.~13 of \citet{Waxman2018}, itself based on the models of \citet{Kasen2017} with the combined data from GW170817/AT2017gfo \citep{Cowperthwaite2017,Drout2017,Villar2017}. 
By combining these quantities and the uncertainties in each model from \citet{Zhu2021}, we infer a disk ejecta mass of $\rm{(2.926-3.645)\times10^{-2}M_\odot}$.
Returning to our assumption that this ejecta mass corresponds to 30-40\% of the original disk mass, this implies a disk mass range of $\rm{(7.32-12.2)\times10^{-2}M_\odot}$.

\section{Equations of State}\label{sec:EOS}

\begin{figure}
    \centering
    \includegraphics[scale=0.7]{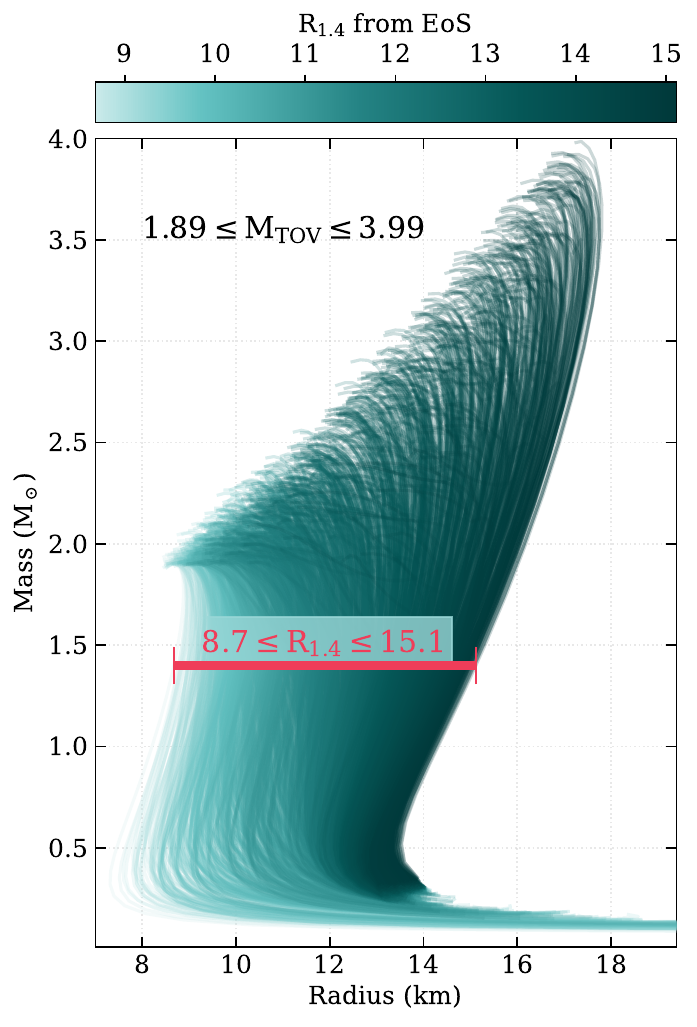}
    \caption{ Mass radius curves for EOSs. The total sample of 2000 EOSs results in a range of $8.7 \leq \rm{R_{1.4}} \leq 15.1$ and $1.89 \leq \rm{M_{TOV}} \leq 3.99$.}
    \label{fig:allowed}
\end{figure}

We use the family of EOSs presented in \citet{Capano2020}.
The details of the construction of this family of EOSs are included in the original publication; we include a summary of the methods used for convenience. 

The construction of our EOS sample begins with microscopic quantum Monte Carlo calculations for the neutron-matter EOS based on two nuclear Hamiltonians from a $\chi$EFT up to $\rm{2 n_{sat}}$, where $\rm{n_{sat}}$ is the nuclear saturation density.
The employed interactions were  fit to nucleon-nucleon scattering data, the $\alpha$-particle binding energy, and neutron-alpha scattering properties~\citep{Tews2018}. 
The neutron-matter EOS  was then extended to $\beta$-equilibrium, and a crust was added in order to obtain neutron star EOSs. 
The high-mass neutron star regime was accessed by computing the speed of sound, $\rm{c_s}$, up to either $\rm{n_{sat}}$ or $\rm{2n_{sat}}$ for the microscopic calculations, then performing a six-point extension of the speed of sound calculation up to $\rm{12n_{sat}}$, with the constraint that $0 < c_s < c$. 
This procedure is carried out for both Hamiltonians for $\sim10,000$ EOSs.
Here, we use results up to $\rm{n_{sat}}$ and solve the Tolman-Oppenheimer-Volkoff (TOV) equations \citep{Oppenheimer1939,Tolman1939} to obtain solutions for the neutron star mass-radius relation for each EOS, and exclude EOSs with maximum masses below 1.9$M_\odot$.
This initial data set is further reduced to 2000 EOSs selected such that the prior on the radius of a 1.4$M_\odot$ NS ($\rm{R_{1.4}}$) is roughly uniform.
Across the resulting EOSs, whose mass-radius curves are shown in Figure~\ref{fig:allowed}, the maximum TOV mass is 3.99 $\rm{M_\odot}$, and the radius of a 1.4$\rm{M_\odot}$ neutron star lies between 8.7 and 15.1 km.

\section{Kilonova  Constraints on Nuclear EOS}\label{sec:EOSconnect}

\begin{figure}[t]
    \centering
    \includegraphics[scale=0.5]{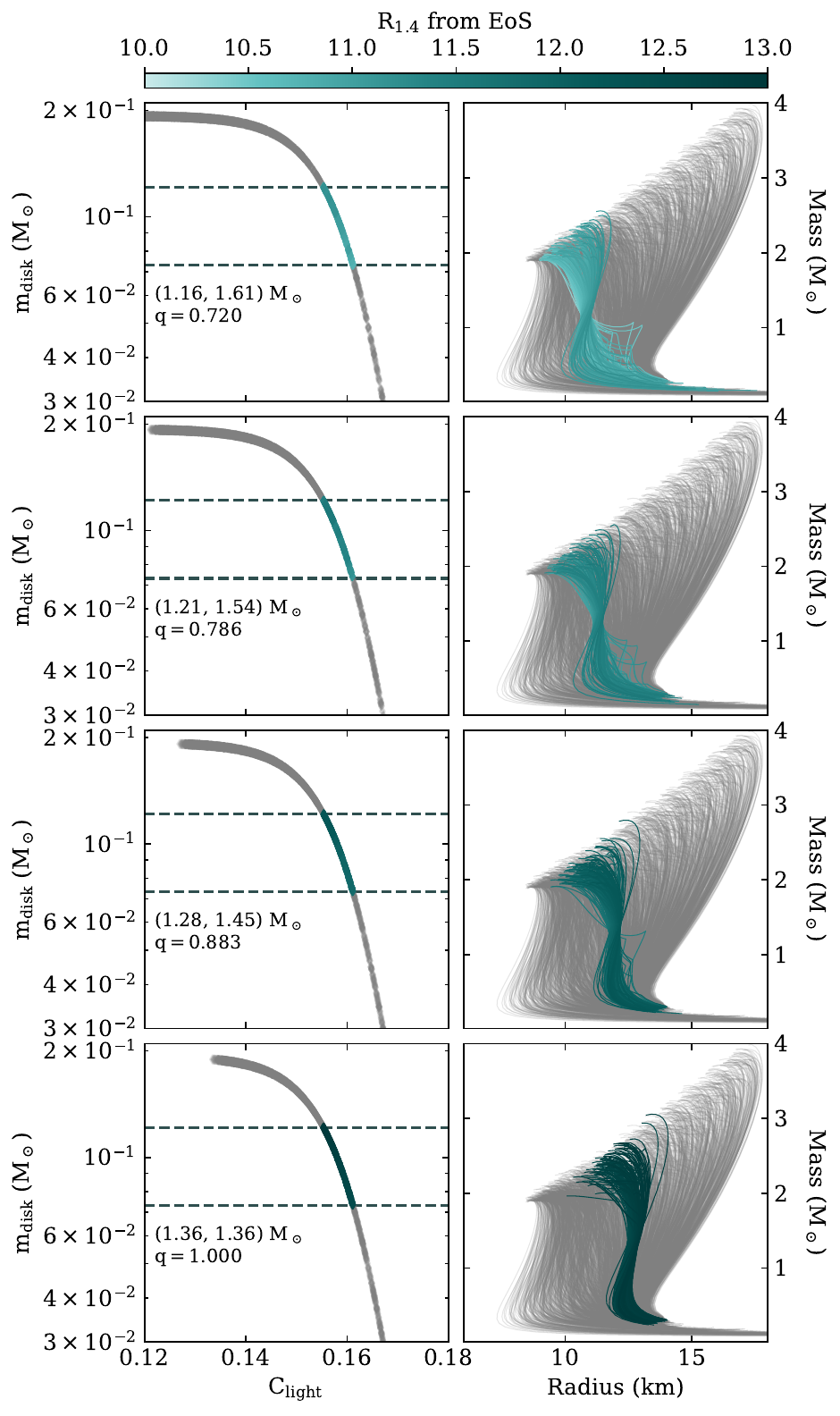}
    \caption{\textit{Left:} disk masses as a function of lightest neutron star compactness, $\rm{C_{light}}$, for four binaries. \textit{Right:} mass-radius curves for allowed EOSs. \textit{Both columns:} for each binary, EOSs that result in disk masses between the inferred values of $0.073-0.122\rm{M_\odot}$ (indicated with dashed lines in the left column) are shown colored according to the $1.4\rm{M_\odot}$ radius ($\rm{R_{1.4}}$), as shown in the color bar. EOSs that do \textit{not} result in masses within the aforementioned range are shown in grey for ease of comparison with Figure~\ref{fig:allowed}.}
    \label{fig:diskmasses_obs}
\end{figure}

The final step in this puzzle is to use our inferred ejecta and kilonova properties to interpret implications for the nuclear EOS. 
For this, we construct four possible binaries consistent with the literature values for the masses involved in GW170817. 
We use the constraint $\rm{1.16 \leq m_{light} \leq 1.36 }$ along with the tight constraint of the chirp mass \citep{Abbott2019}:
\begin{equation}
\mathcal{M} = \frac{\left ( \rm{m_{light}\cdot m_{heavy}}\right )^{3/5}}{\left ( \rm{m_{light} + m_{heavy}}\right )^{1/5} } = 1.186^{+0.001}_{-0.001}\rm{M_\odot}
\end{equation} 
to obtain the mass of the larger companion. 
Although our disk mass formulation in Equation~\eqref{eq:fit} depends solely on the properties of the lighter neutron star, we use both masses to compare our analysis with other formulations in the Appendix.\footnote{We note that the use of updated inferred parameters for the GW170817 binary, such as those in \citet{Breschi2024} would have a small effect on these comparisons, but would not affect our main results unless the individual neutron star mass were also changed.}. 

For each binary, we compute the possible disk masses using Equation~\eqref{eq:fit} for the 2000 EOSs described in Section~\ref{sec:EOS}. 
We show the compactness values (and therefore the EOSs) that result in $7.32\times10^{-2}\rm{M_\odot} \leq \rm{m_{disk}} \leq 12.2\times10^{-2}\rm{M_\odot}$ in Figure~\ref{fig:diskmasses_obs}.

One important consequence is that smaller values of $\rm{m_{light}}$ result in smaller compactness values, therefore favoring softer \footnote{Here, "soft" refers to a particular EOS predicting a smaller radius for a given mass.} EOSs. 
This can be seen in the right column of Figure~\ref{fig:diskmasses_obs}, which shows the mass-radius curves of the allowed EOSs in the left column panels. 
Although we did not use the mass of the larger neutron star in our analysis, the aforementioned tightly constrained chirp mass for GW170817 implies that a neutron star with smaller $\rm{m_{light}}$ will have a larger companion for the same chirp mass. 
Thus it can be interpreted that a softer EOS is also favored for more asymmetric binaries. 

Across all four sample binaries, one of the major outcomes of our analysis is a constraint on the 1.4$\rm{M_\odot}$ radius that characterizes the EOSs with which we work. 
Overall, the allowed EOSs (colored in shades of blue in Figure~\ref{fig:diskmasses_obs}) predicted values of $10.30 \leq \rm{R_{1.4}} \leq 13$ km and $\rm{M_{TOV} \leq 3.06M_\odot}$.
In Appendix~\ref{sec:app_comparison}, we compare the resulting values of $\rm{R_{1.4}}$ and $\rm{M_{TOV}}$ when obtained using the literature fits described in Section \ref{sec:disk}.
Note that observations do not constrain the radii of neutron stars at low masses. 
Hence, very stiff EOS at low densities, leading to large radii, remain valid if they dramatically soften before the observable neutron-star mass regime. 
These EOSs have strong first-order phase transitions, leading to strong softening and hence, jumps in the mass-radius curve that can be observed in Figure~\ref{fig:diskmasses_obs}.

\section{Summary and Discussion}\label{sec:discussion}

In this work, we took a closer look at the complex interplay between the EOS, postmerger accretion-disk evolution, and kilonova observations.
We discussed some of the limitations and implications of different sources of uncertainty, and reevaluated a number of existing fits from the literature that serve to predict a disk mass for a neutron star binary from EOS properties.
Our fit used results from NR simulations, from 11 different sources, for a total of 112 data points. 
We find that the compactness of the lighter binary component is the best indicator of $\rm{m_{ disk}}$.
We propose that the functional form of Equation~(\ref{eq:fit}) performs slightly better than others, with an rms of $0.042$. 
Our formula is limited by the availability of simulation data, which exists for more symmetric binaries. 
The average mass ratio of our data sample was 0.91 with 58 of 112 simulations being equal-mass binaries.
We expect that the inclusion of more data points, especially those from more asymmetric binaries, will result in better fits in the future. 

We then used 3DGRMHD simulations to inform the fraction of disk mass ejected after a binary neutron star event and found a mass fraction of $30-40$\%. 
Using the assumption that the ejecta are entirely responsible for the red component of a kilonova, we then used the observation of $\rm{L_{peak}}$ and $\rm{t_{peak}}$ to estimate ejecta and disk masses.
To first order, the incorporation of a contribution to this luminosity from the dynamical ejecta would imply that less than 100\% of the peak luminosity is attributed to ejecta from the postmerger system. Given that there is no obvious link between disk mass and the \textit{percentage} of the disk that gets ejected, this in turn implies a smaller disk mass. From \fref{fig:diskmasses_obs}, this would push the allowed values of compactness towards higher values, therefore possibly favoring softer EOSs.

Finally, we connected these values backwards to constrain the nuclear EOS, resulting in a prediction of $10.30 \leq \rm{R_{1.4}} \leq 13$ km and $\rm{M_{TOV} \leq 3.06M_\odot}$. 
It is apparent that the values we obtain result in error bars that are larger, but not entirely inconsistent, with other literature values.
We consider, for example, the results of \citet{Bauswein2017}, who make predictions regarding the 1.6$\rm{M_\odot}$ radius guided by fairly conservative assumptions about the properties of GW170817. 
Their analysis was driven mostly by the constraint provided by the EOS-dependent threshold mass, $\rm{M_{thres}}$- the same threshold mass used in the fits of \citet{Coughlin2018} and \citet{Dietrich2020}.
It was concluded that the minimum radius of a 1.6$\rm{M_\odot}$ neutron star must be $10.68^{+0.15}_{-0.04}$ km. 
Similarly, both \citet{Koppel2019} and \citet{Kashyap2022} obtained a similar constraint of $\rm{R_{1.6} \geq 10.90}$ km.
Had we operated only under the assumption of an equal-mass binary, with each mass being 1.36$\rm{M_\odot}$ (which is the case for the bottom row of Figure~\ref{fig:diskmasses_obs}), the allowed EOSs from our sample result in a similar prediction of $\rm{R_{1.6}} \geq 10.94$. 

Our analysis leads to results based on simple statistics informed by a single event.
It is therefore not surprising that the width of our EOS error bars is larger than, for example, the results published in \citet{Koehn2025}, which are based on Bayesian statistics of a wide variety of astronomical constraints. 
However, even their analysis of the combined gravitational wave $+$ kilonova $+$ gamma-ray burst data leads to an estimate of $\rm{R_{1.4}} = 12.19^{+0.71}_{-0.63}$ km, which is in good agreement with the range of radii we obtain. 
It should be emphasized that the results of \citet{Koehn2025} are given in terms of 90\% confidence levels; ours are meant to simply illustrate possible values given a detailed look at specific aspects of the analysis. 
Similarly, the combined gravitational wave + kilonova + pulsar analysis presented in \citet{Breschi2024} leads to estimates of $\rm{R_{1.4} = 12.30^{+0.81}_{-0.56}\ (13.20^{+0.91}_{-0.90})}$ km and $\rm{M_{TOV} = 2.28 ^{+0.21}_{-0.17}}\ (2.32^{+0.30}_{-0.19})\ M_\odot$, with parentheses indicating the use of different analysis results of the pulsar J0030+0451. Though the uncertainty on our results is larger, they are not at odds with these narrower constraints.

We highlight the importance of the underlying physics that is often overlooked in favor of fit formulae used to obtain point estimates of, for example, the remnant accretion-disk mass or the ejecta mass from that disk.
By propagating this uncertainty through the many degeneracies in a full inference, we hope to motivate studies aimed at probing these different physical problems. 
We further hope to incorporate more robust statistical methods in future work, making our approach more generally applicable to frameworks like the one described in \citet{Pang2023}.
We look forward to our proposed fitting model being put to the test with new simulation data.

\section{Acknowledgements}
We thank M. Bulla and T. Dietrich for their helpful comments during the preparation of the manuscript.

This document has been approved for unlimited release, assigned LA-UR-24-24836.
K.A.L. and I.T. were supported by the Laboratory Directed Research and Development program of Los Alamos National Laboratory under project number 20230315ER.
K.A.L., M.R.M., and I.T. also acknowledge support from the Laboratory Directed Research and Development program of Los Alamos National Laboratory under project number 20230052ER.
I.T. was also supported by the U.S. Department of Energy, Office of Science, Office of Nuclear Physics, under contract No.~DE-AC52-06NA25396, and by the U.S. Department of Energy, Office of Science, Office of Advanced Scientific Computing Research, Scientific Discovery through Advanced Computing (SciDAC) NUCLEI program.
R.S. acknowledges support from the Nuclear Physics from Multi-Messenger Mergers (NP3M) Focused Research Hub which is funded by the National Science Foundation under Grant Number 21-16686, and by the Laboratory Directed Research and Development program of Los Alamos National Laboratory under project number 20220541ECR.
K.A.L. and M.R.M acknowledge support from the Directed Asymmetric Network Graphs for Research (DANGR) initiative at Los Alamos. 
J.M.M. acknowledges support from LDRD project 20220564ECR.
We gratefully acknowledge the support of the Center for Nonlinear Studies (CNLS) at Los Alamos National Laboratory for this work. 
Los Alamos National Laboratory is operated by Triad National Security, LLC, for the National Nuclear Security Administration of U.S.\ Department of Energy (Contract No.\ 89233218CNA000001). G.C.M acknowledges support from the NSF (N3AS PFC) grant No. PHY-2020275, as well as from U.S. DOE contract Nos. DE-FG0202ER41216 and DE-SC00268442 (ENAF), as well as by LA22-ML-DE-FOA-2440.  This work is performed in part under the auspices of the U.S. Department of Energy by Lawrence Livermore National Laboratory under Contract DE-AC52-107NA27344, with support from LDRD project 24-ERD-023.

%%%%%%%%%%%%%%%%%%%%%%%%%%%%%%%%%%%%%%%%%%%%%%%%%%%%%%%%%%%        Appendix        %%%%%%%%%%%%%%%%%%%%%%%%%%%%%%%%%%%%%%%%%%%%%%%%%%%%%%%%%%%%
\newpage
%\begin{comment}
\appendix
\section{Disk Mass Fit Formulae\label{sec:app_Fits}}
In this appendix, we describe some of the proposed methods for estimating the resulting disk mass from an NSM based on fits to data from NR simulations. 

\subsection{Binary Tidal Deformability}

The tidal deformability of a binary system depends on the tidal deformability of the individual binary components (from \citet{De2018}): 

\begin{equation}
    \rm{\widetilde{\Lambda} = \frac{16}{13}\frac{(12q+1)\Lambda_{heavy} + (12+q)q^4\Lambda_{light}}{(1+q)^5}},\ \rm{where}
\end{equation}
\begin{equation}
    \rm{q=\frac{m_{light}}{m_{heavy}}\leq 1,}\ \rm{and}
\end{equation}
\begin{equation}
    \rm{\Lambda_{light,heavy} = \frac{2}{3}k_2\left ( \frac{R_{light,heavy}c^2}{Gm_{light,heavy}}\right )^5},
\end{equation}
where the tidal Love number, $\rm{k_2}$, depends on the neutron star's mass and EOS.

\citet[][R18]{Radice2018} evaluated a grid of 35 numerical relativity (NR) simulations with 4 different EOSs. 
From the results of these calculations, they propose a best-fit formula for the disk mass that depends on the tidal deformability, $\widetilde{\Lambda}$, of the binary:
\begin{equation}\label{eq:lambda}
    \rm{\frac{m_{disk}}{M_\odot} = {\rm max}\left\{ 10 ^{-3}, \alpha + \beta \tanh\left(\frac{\widetilde{\Lambda} - \gamma}{\delta}\right)\right\}},
\end{equation}
with 
$\rm{\alpha=0.084}$, 
$\rm{\beta=0.127}$, 
$\rm{\gamma=567.1}$, and 
$\rm{\delta=405.14}$. 

\subsection{Prompt Collapse Threshold Mass}
\citet{Coughlin2019} take the NR simulations from R18 and highlight that the lifetime of the postmerger remnant is related to the stability of said remnant, and this lifetime is strongly correlated with the resulting disk mass. 
The remnant lifetime prior to collapse is governed in large part by the ratio of the binary mass to the threshold mass (above which there is prompt collapse to a black hole).
The threshold mass can be computed following \citet{Bauswein2013b}:
\begin{equation}
    \rm{M_{thr} = \left( -jC^*_{1.6}+a \right) M_{TOV},\ where}
\end{equation}
\begin{equation}
    \rm{C^*_{1.6} = \frac{GM_{TOV}}{c^2R_{1.6}}},
\end{equation}
and best-fit parameters $\rm{j=3.606}$ and $\rm{a=2.380}$.

Based on the NR simulations from R18 and the correlation of the binary threshold mass, $\rm{M_{thr}}$, with the resulting disk mass, \citet{Coughlin2019} propose the following relation:
\begin{equation}\label{eq:thresh}
    \rm{log_{10} \left( \frac{m_{disk}}{M_\odot}\right) =} \\ {\rm{max}} \left\{ -3, \rm{a}\left( \rm{1+b}\tanh{\rm{\frac{c-M_{tot}/M_{thr}}{d}}}\right)\right\},
\end{equation}%\end{multline}
where $\rm{M_{tot}}$ is the total binary mass, while $\rm{a=-31.335}$, $\rm{b=-0.9760}$, $\rm{c=1.0474}$, and $\rm{d=0.05957}$ are the best-fit parameters. 

However subsequent work (D20) compares results from 73 NR simulations performed by various groups resulting in a modified formulation of the resulting disk mass that incorporates a dependence on the binary mass ratio by modifying the parameters such that

\begin{align}
    \rm{a} &= \rm{a_0 + \delta a\cdot\xi} \\ \rm{b} &= \rm{b_0 + \delta b\cdot\xi},
\end{align}
where the parameter $\xi$ is given by:
\begin{equation}
     \rm{\xi = \frac{1}{2}\tanh \left( \beta\left( \hat{q}-\hat{q}_{trans}\right)\right)}.
\end{equation}
Here $\rm{\hat{q} = m_{light}/m_{heavy} \leq 1}$ is the inverse binary mass ratio; $\rm{\hat{q}_{trans}}$ and $\beta$ are free parameters. D20 report best-fit parameters 
$\rm{a_0 = -1.581}$,
$\rm{\delta a = -2.439}$,
$\rm{b_0 = -0.538}$,
$\rm{\delta b = -0.406}$,
$\rm{c = 0.953}$,
$\rm{d = 0.0417}$,
$\rm{\beta = 3.910}$, and 
$\rm{\hat{q}_{trans}=0.900}$.

\subsection{Lightest NS Compactness}
Additional efforts to continue to improve upon the results from R18 and \citet{Coughlin2019} were made by KF20 by incorporating disk masses from an additional 22 NR simulations from \citet{Kiuchi2019} that included asymmetric binary mass ratios. 
These efforts resulted in a formulation dependent on the compactness parameter of the lighter of the two neutron stars, $\rm{C_{light}}$:
\begin{equation}\label{eq:com}
    \rm{m_{disk} = m_{light}\cdot max\left\{ 5\times10^4, \left( aC_{light} + c \right)^d\right\}},
\end{equation}
with best-fit parameters
$\rm{a = -8.1324}$,
$\rm{c = 1.4820}$, and
$\rm{d = 1.7784}$.

\subsection{Comparison to Other Fits}\label{sec:app_comparison}
We include in Table~\ref{tab:results_compare} a comparison of our overall result, obtained starting from Equation~\eqref{eq:fit}, to those we would have obtained had we used the methods described in this Appendix.

\begin{table*}[h!]
    \centering
    \begin{tabular}{|c||c|c||c|c|c|}
    \hline
        \textbf{Binary} & \textbf{Value} & \textbf{Our Results} & \textbf{KF20} & \textbf{R18} & \textbf{D20}\\
    \hline
        \multirow{2}{8em}{1.16 $\rm{M_\odot}$, 1.61 $\rm{M_\odot}$} & $\rm{R_{1.4}}$ (km) & 10.30-11.23 & 10.68-11.70 & 11.46-13.02 & 10.26-11.73\\
        & $\rm{M_{TOV}\ (M_\odot)}$ & 1.90-2.56 & 1.89-2.56 & 1.91-2.94 & 1.89-2.36\\
        \hline
        \multirow{2}{8em}{1.21 $\rm{M_\odot}$, 1.54 $\rm{M_\odot}$} & $\rm{R_{1.4}}$ (km) & 10.97-11.64 & 10.87-12.19 & 11.46-12.90 & 10.64-12.05\\
        & $\rm{M_{TOV}\ (M_\odot)}$ & 1.90-2.56 & 1.89-2.80 & 1.91-2.94 & 1.91-2.36\\
        \hline
        \multirow{2}{8em}{1.28 $\rm{M_\odot}$, 1.45 $\rm{M_\odot}$} & $\rm{R_{1.4}}$ (km) & 11.46-12.26 & 11.46-12.72 & 11.55-12.85 & 10.77-12.49\\
        & $\rm{M_{TOV}\ (M_\odot)}$ & 1.90-2.80 & 1.90-2.94 & 1.91-2.94 & 1.90-2.43\\
        \hline
        \multirow{2}{8em}{1.36 $\rm{M_\odot}$, 1.36 $\rm{M_\odot}$} & $\rm{R_{1.4}}$ (km) & 12.50-12.96 & 12.68-13.38 & 12.16-12.82 & 11.01-12.54\\
        & $\rm{M_{TOV}\ (M_\odot)}$ & 1.96-3.06 & 1.92-3.06 & 1.91-2.94 & 1.96-2.56\\
        \hline
        \multirow{2}{8em}{Overall} & $\rm{R_{1.4}}$ & 10.30-12.96 & 10.68-13.38 & 11.46-13.02 & 10.26-12.54\\
        & $\rm{M_{TOV}\ (M_\odot)}$ & 1.90-3.06 & 1.89-3.06 & 1.90-2.94 & 1.89-2.56\\
        \hline
    \end{tabular}
    \caption{Comparison of EOS values using other fits from the literature, as described in this Appendix.}
    \label{tab:results_compare}
\end{table*}

\bibliography{ref}
\end{document}